numbers=left                            %%
\documentclass[pdflatex,sn-basic,a4paper]{sn-jnl}% Basic Springer Nature Reference Style/Chemistry Reference Style

\usepackage{fixme}
\usepackage{listings}
\usepackage{acronym}

\jyear{2021}%

\theoremstyle{thmstyletwo}%

\theoremstyle{thmstylethree}%

\lstdefinestyle{defaultStyle}{
basicstyle=\small\ttfamily,
keywordstyle=\color{black}\underbar,
numbers=left
}

\begin{document}

\nocite{*}

\title{Using \acs{DevOps} Toolchains in Agile Model-Driven Engineering}


\author*[1]{\fnm{Jörn Guy} \sur{Süß}}\email{joern.guy.suess@codebots.com}

\author[1]{\fnm{Samantha} \sur{Swift}}\email{samantha.swift@codebots.com}
\equalcont{These authors contributed equally to this work.}

\author[1]{\fnm{Eban} \sur{Escott} \dgr{PhD}}\email{eban.escott@codebots.com}
\equalcont{These authors contributed equally to this work.}

\affil*[1]{\orgdiv{Research and Development}, \orgname{Codebots Pty. Ltd}, \orgaddress{\street{55 Railway Terrace}, \city{Milton}, \postcode{4064}, \state{Queensland}, \country{Australia}}}

\abstract{For \ac{MDE} to become more Agile the community needs to embrace \ac{DevOps} practices. One of the core practices of DevOps is the use of pipelines to enable CI/CD to make teams more Agile and break down the barriers between development and operations with faster deployments.
Current MDE tooling is not designed at its core to participate in DevOps pipelines. Consequently this makes the adoption of MDE in industry more difficult. In this article, we cover an industrial experience report describing how we enabled our pipelines using DevOps and MDE.}
%https://www.acm.org/publications/class-2012

%%\pacs[ACM Classification]{D8, H51}

\keywords{DevOps, CI/CD, Ant, EMF, Eclipse, Agile, Model-Driven Engineering, MDE}



\maketitle

%\nocite{*}

\section{Introduction}

\ac{MDE} has a strong representation in teaching and research contexts, while in industry, it mainly reaches success in areas with firmly established domain languages (\cite{DBLP:conf/models/WhittleHRBH13}). For MDE to be more widely accepted into Agile teams, more modern development practices like DevOps must be considered.

To facilitate further adoption, we must address the current barriers to adoption and enable Agile MDE using DevOps. Modern DevOps practices combine tools into pipelines to improve quality and increase release cycle speed. This article provides an experience report and architecture description reflecting our companies journey to make \ac{MDE} Agile by using \ac{DevOps} and toolchains as a vehicle.

When we began this project, we found that industrial practice and requirements in our company were quite different to those for teaching and research: Artifacts need to be rigorously built, tested, integrated and deployed in an efficient and fast manner. These requirements are also at the heart of the Agile approach to software production that we practice. It is worth pointing out that Agile approaches have always been strongly connected to automation or mechanisation to support a fast turn-around, and provide feedback on quality in small increments (\cite{fowler05integration}).

\begin{figure}[ht]
\centering
\includegraphics[width=\textwidth]{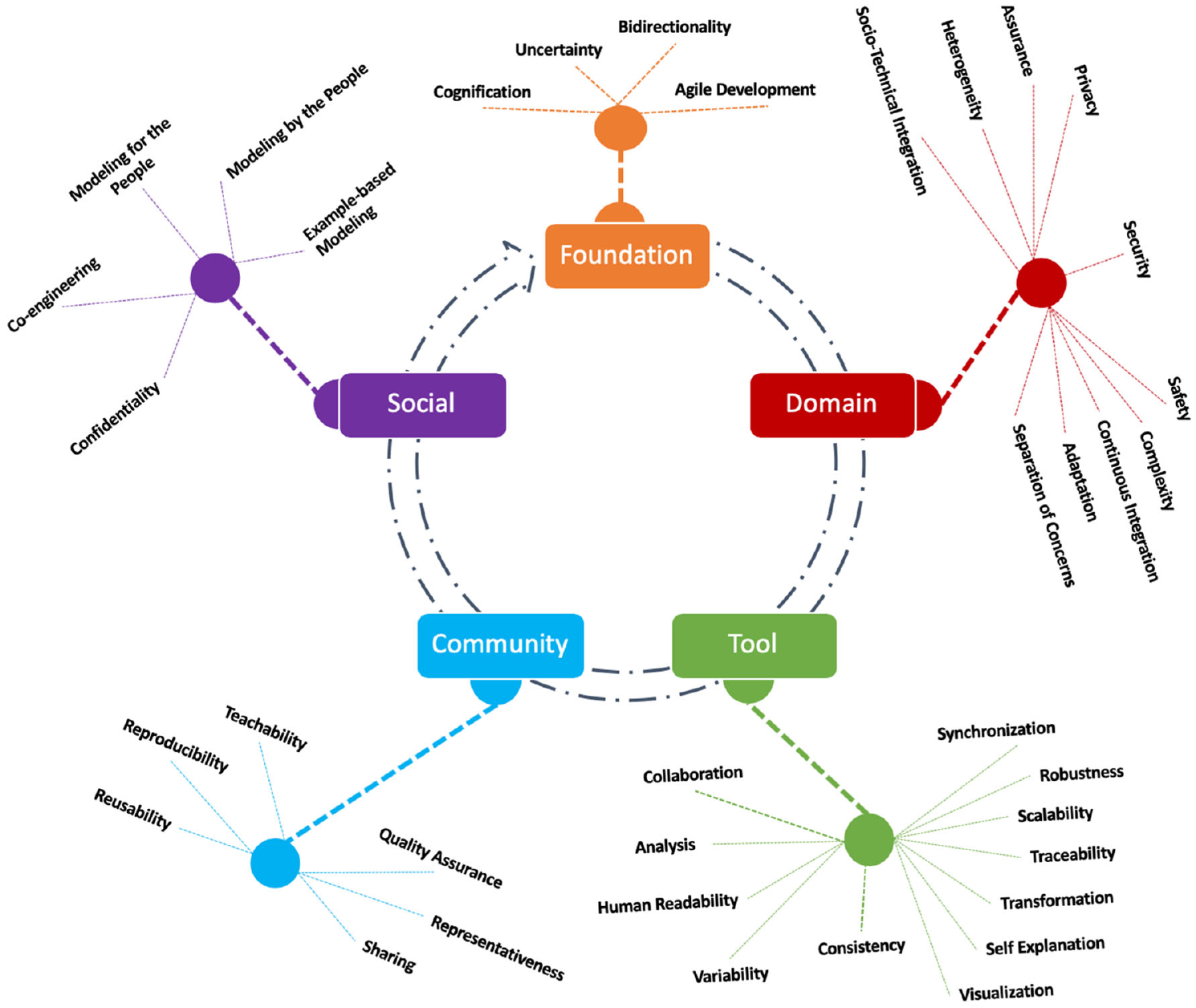}
\caption{\ac{MDE} Grand Challenges from~\cite{DBLP:journals/sosym/BucchiaroneCPP20}}
\label{MDE_CHALLENGES}
\end{figure}

In the model-driven community, discussions about the causes of the lack of general industrial adoption have occurred for a long time (\cite{DBLP:conf/models/StraetenMB08,DBLP:journals/sosym/BucchiaroneCPP20}).
However, while the community seems to look at the gains and successes of Agile approaches with some envy, it seems to pay less attention to the technical foundations that enable it: \autoref{MDE_CHALLENGES} shows a taxonomy of what community consensus perceived as the Grand Challenges to the adoption of \ac{MDE} in 2020: \ac{CI} receives a marginal mention;  \ac{CD} and \ac{DevOps}, the practices and operations that drive Agile processes at large these days, are not present at all. Especially \ac{DevOps} - the programming of the process that produces and deploys software quickly, efficiently and with timely pertinent feedback to stakeholders (\cite{DBLP:journals/usenix-login/Debois11}) - is pivotal to a contemporary software developers' workflow.

From our perspective as a software development company, current \ac{MDE} tooling does not meet the technical requirements to deliver Agile operations in industrial practice. This article posits that if the \ac{MDE} community wants to have practical industrial relevance, and operate in an Agile manner, it must consider \ac{DevOps} and \ac{CI}/\ac{CD} a first-class concern. It must also provide for horizontal reuse, allowing \emph{all} \ac{MDE} tooling to be reused within programmed workflows.
 
This prompts the question, why it is challenging to horizontally integrate and reuse existing \ac{MDE} tooling and how one would meet that challenge? In the remainder of this section, we investigate the impediments we have experienced in detail. In \autoref{sec:architecture} we describe how our architecture is designed to meet the challenges using a combination of tried off-the-shelve components and minimal additions.

\subsection{Challenges of Using Eclipse}

The common foundation for most \ac{MDE} offerings is the \ac{EMF}; \ac{EMF} is a light-weight Java-based descendant of the \ac{MOF}, the original meta-modelling service embedded in the complex \ac{CORBA}. The dependency on \ac{EMF} implies that most tool-sets are also expressed within the Eclipse \ac{RCP} or have some dependency on it.

\ac{RCP} is a challenging platform to reliably write software for. It consists of components, has \href{https://newsroom.eclipse.org/news/announcements/eclipse-ide-turns-20}{evolved over 20 years}, and substantially relies on frameworks and declarative mechanisms. Compounded by thin documentation and challenging access to source code, Eclipse is a major hurdle in the adoption of \ac{EMF}-based software.

At the same time, \ac{RCP} is feature-rich and its \ac{UI} facilities are directly linked with generated artifacts that \ac{EMF} produces. This more often than not leads to close coupling of the model-driven tools to the Eclipse \ac{UI}. For research demonstrators or teaching use, this does not cause a problem and even provides an advantage to productivity due to tight and direct reuse.

If we want to retain access to functionality that is currently implemented, we need to accept Eclipse \ac{RCP} as a platform. Most \ac{MDE} toolkits offer some automation. But how do we co-ordinate the execution of \emph{all \ac{MDE} toolkit functionality} within a single platform? If we would be able to show this to be practical, we could commercially combine high-quality existing offerings to create exciting new solutions. Showing this functionality might eventually entice \ac{MDE} tool developers to invest in a separation of their user interface from their services functions and models, ultimately leading to a pool of Agile \ac{DevOps}-capable \ac{MDE} toolchain components.

\subsection{Ant in Eclipse: A minimal vehicle for \ac{MDE} CI/CD}
\label{sec:ant_as_vehicle}

Historically, Eclipse has an interactive, declarative build system. It consists of a file-system that detects changes, and \href{https://help.eclipse.org/latest/topic/org.eclipse.platform.doc.isv/reference/extension-points/org_eclipse_core_resources_builders.html}{Builders that respond to them and update derivative files} and views in various ways. This mode of development was also initially used for the development of Eclipse components themselves. However, it quickly became clear that this approach did not scale.

Consequently Eclipse's \ac{PDE} was extended to be controlled by the \href{https://ant.apache.org/}{Ant build system}. At that time, Ant was the standard build facility for Java-associated projects. Since \ac{PDE} made use of Eclipse internals, an Eclipse host had to be running so the Ant build tasks for Eclipse build automation could use the underlying framework-based facilities. This resulted in the design of a headless Eclipse host application known as the \href{https://help.eclipse.org/latest/topic/org.eclipse.platform.doc.isv/guide/ant_running_buildfiles_programmatically.htm}{AntRunner} that thinly wraps around the Ant execution engine.

Since the following discussion heavily depends on an understanding of Ant, we will give a limited characterisation here. Ant is a build system that exposes Java-written functionalities in an XML document using elements called \href{https://ant.apache.org/manual/using.html#tasks}{Tasks}. Tasks are implemented as Java classes that are dynamically loaded at runtime. Task instances are grouped into sequences of functionality called \href{https://ant.apache.org/manual/using.html#targets}{Targets} that represent desirable outcomes to the end-user. Targets may depend on each other. This makes Ant build specifications directed multi-graphs. The user selects a Target and the engine traverses and executes prerequisite Targets depth-first. The Ant runtime has a \href{https://ant.apache.org/manual/using.html#references}{map-like storage for global state} and a facility intended for asserting facts in so-called \href{https://ant.apache.org/manual/using.html#properties}{Properties}. Task instances can interact with the global state and make assertions, affecting execution. Tasks may \href{https://ant.apache.org/manual/api/org/apache/tools/ant/Project.html#executeTarget(java.lang.String)}{dynamically invoke targets}. This property, combined with the global mutable state, means that Ant is effectively Turing-complete.

To run an Ant build a user minimally has to provide an Ant build file, any initial properties, the list of Targets to build and implementations for all Tasks that will be invoked in the build execution. Further, Ant has an input handling mechanism that can synchronously obtain information from a user at runtime. \autoref{lst_ant_example} shows an \href{https://ant.apache.org/manual/using.html#example}{example build file from the Ant online documentation} to exemplify these concepts.

\begin{lstlisting}[language=Ant, label=lst_ant_example, caption=An example of an Ant build file.]
<project name="MyProject" default="dist" basedir=".">

  <description>
    simple example build file
  </description>
  
  <!-- set global properties for this build -->
  <property name="src" location="src"/>
  <property name="build" location="build"/>
  <property name="dist" location="dist"/>

  <target name="init">
  
    <!-- Create the time stamp -->
    <tstamp/>
    
    <!-- Create the build directory structure used by compile -->
    <mkdir dir="${build}"/>
    
  </target>

  <target name="compile" depends="init"
          description="compile the source">
          
    <!-- Compile the Java code from ${src} into ${build} -->
    <javac srcdir="${src}" destdir="${build}"/>
  </target>

  <target name="dist" depends="compile"
          description="generate the distribution">
          
    <!-- Create the distribution directory -->
    <mkdir dir="${dist}/lib"/>

    <!-- Put everything in ${build} -->
    <!-- into the MyProject-${DSTAMP}.jar file -->
    <jar jarfile="${dist}/lib/MyProject-${DSTAMP}.jar"
         basedir="${build}"/>
  </target>

  <target name="clean"
          description="clean up">
          
    <!-- Delete the ${build} and ${dist} directory trees -->
    <delete dir="${build}"/>
    <delete dir="${dist}"/>
  </target>
\end{lstlisting}

Most \ac{MDE} tools delivered as Eclipse features include tasks for the AntRunner. They generally use the interactive Ant execution facility built into the IDE itself to provide a simple build process scripting facility scoped to the tool at hand. Consequently, AntRunner is a minimal point of integration for a large number of \ac{EMF}/\ac{MDE} tool-sets that otherwise would require custom-coordination written in Java. Examples of components that expose through Ant include \href{https://www.eclipse.org/epsilon/doc/workflow/}{Epsilon},  \href{https://help.eclipse.org/latest/index.jsp?topic=\%2Forg.eclipse.emf.mwe.doc\%2Fhelp\%2Fworkflow_reference_workflow_configuration.html&anchor=workflow_reference_starting_from_your_own_code}{Model Workflow Engine 2 (Xtext)},  \href{https://help.eclipse.org/latest/topic/org.eclipse.m2m.atl.doc/guide/user/The-ATL-Tools.html?cp=6_1_4_2#ATL_ant_tasks}{Atlas Transformation Language}, and the  \href{https://download.eclipse.org/modeling/emf/emf/javadoc/2.5.0/org/eclipse/emf/ant/taskdefs/EMFTask.html}{Eclipse Modelling Framework Core}. In addition, \href{http://www.ant4eclipse.org/}{Eclipse Launch Configurations can be started from Ant}, which enables the use of tooling provided by \href{https://help.eclipse.org/latest/topic/org.eclipse.modisco.infrastructure.doc/mediawiki/workflow/user.html}{Modisco's Workflow Component}. Finally, there is a fair supply of \href{https://ant.apache.org/manual/tasklist.html}{other software engineering Ant tasks} ranging from specific system control to general purpose language facilities.

\subsection{Reusing Ant Automation}

However, components integrated via the AntRunner suffer from four main issues that make use of \ac{MDE} in a \ac{CI}/\ac{CD} environment challenging: The components frequently depend on Eclipse \ac{UI} features,  perform slowly due to interpreted and reflective languages, have issues with parallelisation, and have classpath issues due to Ant's limited isolation. The following paragraphs explain these issues in detail.

As stated, most \ac{EMF}/\ac{MDE} tools are written for interactive use in \ac{RCP} and hence reference the \ac{UI} facilities and the Eclipse project-management file system, known as the workspace. \ac{CI}/\ac{CD} and \ac{DevOps} are inherently using console-based interfaces. Build servers in the cloud do not have graphics cards.  Consequently this mismatch causes issues whenever these Eclipse facilities are required in a headless run, but are not available, because they only exist when the \ac{UI} is active. For example, Epsilon, one of the premier integration kits for MDE, has numerous dependencies from its workflow to the UI.  It is possible to compensate for this issue by using an \href{https://en.wikipedia.org/wiki/Xvfb}{X11 virtual framebuffer}, as long as no interactions are required from the graphical interface, but ultimately this is an unsafe procedure and can lead to unexpected locking behaviour the causes of which are very hard to diagnose.

\ac{MDE} toolsets often use weakly typed languages to execute interpreters that work using reflection, to achieve the flexibility that provides the advantage of \ac{MDE} tooling. Safety and speed compared to compiled languages is generally poor, so tool performance and code maintenance also scale poorly. Epsilon, for example, uses untyped languages that are reflectively interpreted over an abstracted feature-based metamodel. This occasionally leads to unexpected type errors when typed expressions are used, in turn invalidating the use of types as a general practice.

\ac{MDE} toolsets are often written with shared state in mind. This is another side-effect of writing for the Eclipse IDE's UI. It assumes that a single user operates sequentially on a single set of data. Static and global registrations and unsynchronized accesses are common. For \ac{CI} workflows this has the implication that it is generally difficult to accelerate a build execution by running parts of the work in parallel. Experimenting with Epsilon, we found it usually causes a \lstinline{ConcurrentModificationException} when used in an \href{https://ant.apache.org/manual/Tasks/parallel.html}{Ant parallel block}. This compounds the performance issue, as parallelization would usually be a strategy to compensate for low interpretation speed.

One of the weaknesses of Ant itself is that it does not manage tasks' dependencies as components, which \href{https://ant.apache.org/manual/Tasks/typedef.html}{can lead to classpath clashes}. The Eclipse AntRunner however manages dependencies using the Eclipse dependency management system. This allows Ant to be used in a reliable fashion. Hence this weakness mostly affects Tasks that were written without the Eclipse container in mind. Packaging them in Eclipse resolves this problem for such components.

As a company, we require the full scope and breadth of tools available in the \ac{MDE} community to create exciting and novel solutions for our customers. Hence enabling reuse is our overriding architecture goal. To us, this goal is at this time more important then model-specific aspects like those raised in ~\cite{DBLP:conf/models/0001KP20}. We aim to compensate the shortcomings named above through engagement in the community. We hope this will help to shape the discussion. This article is intended as a contribution to this goal.

The following section describes the BB8\footnote{Since our company is based on making coding support systems known as Bots available to the public, and we need to build these internally, we dubbed the system architecture BotBuilder. As it offers twice the variability, speed and access to tools as compared to its predecessors, its version number is $2^3=8$. And we admit a slight nod to Star Wars as well.} architecture we have developed to allow this reuse.

\section{BB8: An Architecture for \ac{MDE} \ac{DevOps}}
\label{sec:architecture}

The following sections describe the architecture of our system from the inside out, giving a developers view (sections \ref{ss:AntRunner}-\ref{ss:AntUnit}) and an operations view (section \ref{ss:P2}- \ref{ss:Kubernetes}).

The developers view starts with our reuse of \href{https://help.eclipse.org/latest/topic/org.eclipse.platform.doc.user/concepts/concepts-antsupport.htm?resultof=\%22\%61\%6e\%74\%22\%20}{Eclipse's Ant Support} (\ref{ss:AntRunner}) to run workflows in \ac{CI}. Section \ref{ss:AntHarness} describes how workflows can be tested using JUnit. Section \ref{ss:AntUnit} shows how Ant users can test workflows without requiring knowledge of Java or Eclipse.

The operations view begins in section \ref{ss:P2} with an explanation of how  components of the workflow can be installed and removed. It addresses how a user can get the desired \ac{MDE} toolsets and other tools for running a specialized workflow. Section \ref{ss:Docker} describes how the workflow runtime can be used everywhere, without installing anything locally and with faithful reproduction of results. Section \ref{ss:CI} describes how we deploy that runtime in the \ac{CI} system provided by Gitlab. Finally \ref{ss:Kubernetes} describes how execution is coordinated within cloud resources.

\begin{figure}[t]
\centering
\includegraphics[width=\textwidth]{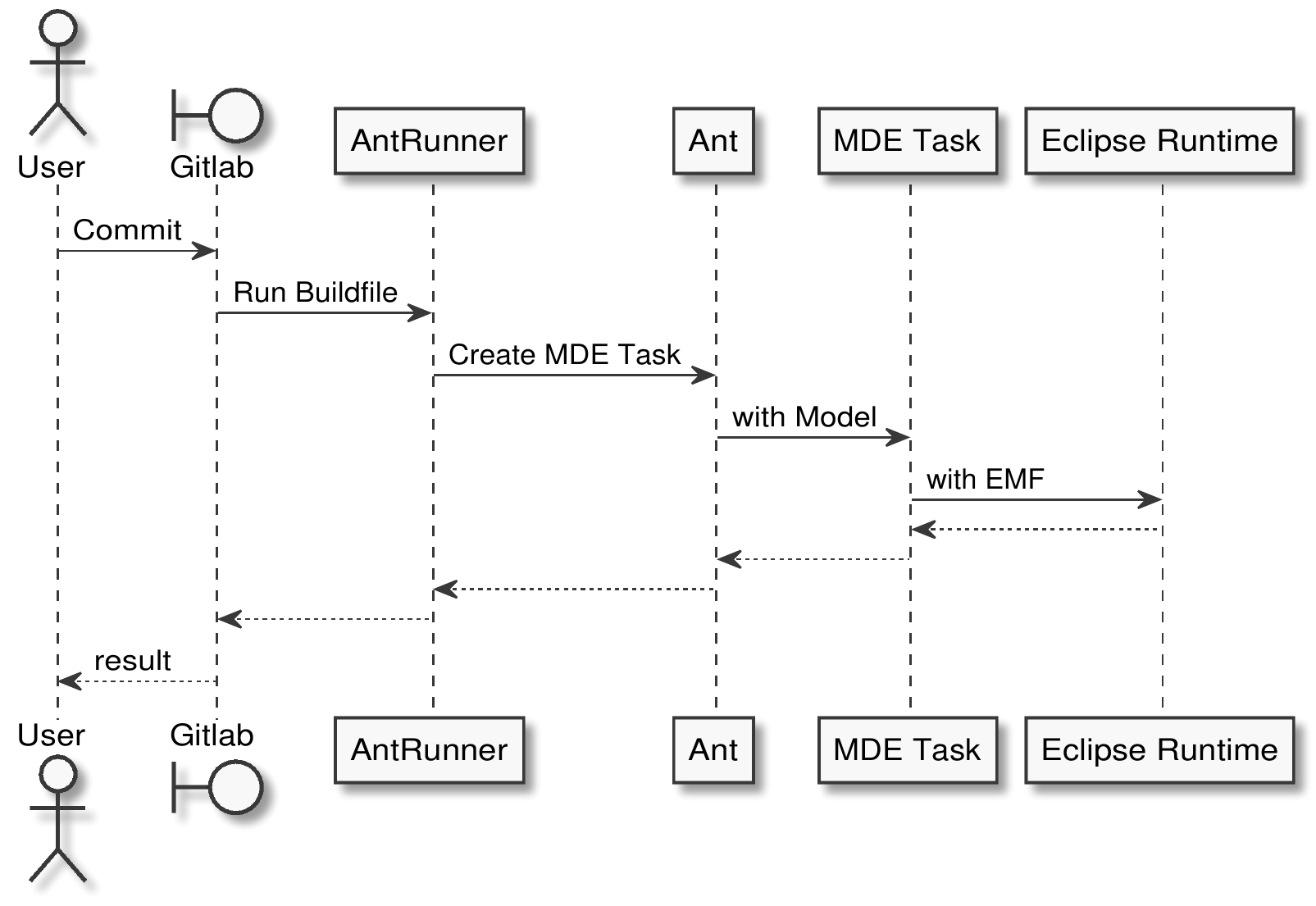}
\caption{A model-driven task runs as the result of a commit to Gitlab CI/CD.}
\label{MDE_CI_CD}
\end{figure}

\subsection{Executing \ac{MDE} Workflows using AntRunner}
\label{ss:AntRunner}

As described in \autoref{sec:ant_as_vehicle}, Eclipse AntRunner is a headless Eclipse application, which is able to execute the Ant build system, while providing access to the Eclipse plugin system and the underlying framework runtime. As noted this allows the reuse of model-driven tooling provided by other projects within the Eclipse scope. It also allows access to all of the \ac{EMF} facilities.

Also, the \href{https://help.eclipse.org/latest/nav/4_2_0}{\ac{PDE}/Build} system used for creating Eclipse components and applications is available in the AntRunner. While \ac{PDE}/Build is not a reliable means to build large-scale Eclipse applications, it is capable of building small applications reasonably. This means that an on-board solution for compiling and packaging model-driven components is in principle available to every workflow without a need to learn additional build facilities like \href{https://projects.eclipse.org/projects/technology.tycho}{Tycho} or \href{https://bnd.bndtools.org}{BND}.

Finally, the Ant build language can express \href{https://ant.apache.org/manual/Tasks/include.html}{includes} and \href{https://ant.apache.org/manual/Tasks/import.html}{imports} of remote build files, and define \href{https://ant.apache.org/manual/targets.html#extension-points}{extension points}, \href{https://ant.apache.org/manual/Tasks/macrodef.html}{macros} and \href{https://ant.apache.org/manual/Tasks/scriptdef.html}{scripts}. Used with care, these features allow for the design of compositional builds.

\subsection{Eclipse-level Testing using AntHarness}
\label{ss:AntHarness}

If we are using Ant to integrate workflows, we will frequently create Ant Tasks to provide secondary functionality and hence we will be developing and debugging a lot of them. Ant Task classes themselves can be unit-tested with relative ease, but interactions with Eclipse's facilities can be complex.

\begin{figure}[ht]
\centering
\includegraphics[width=\textwidth]{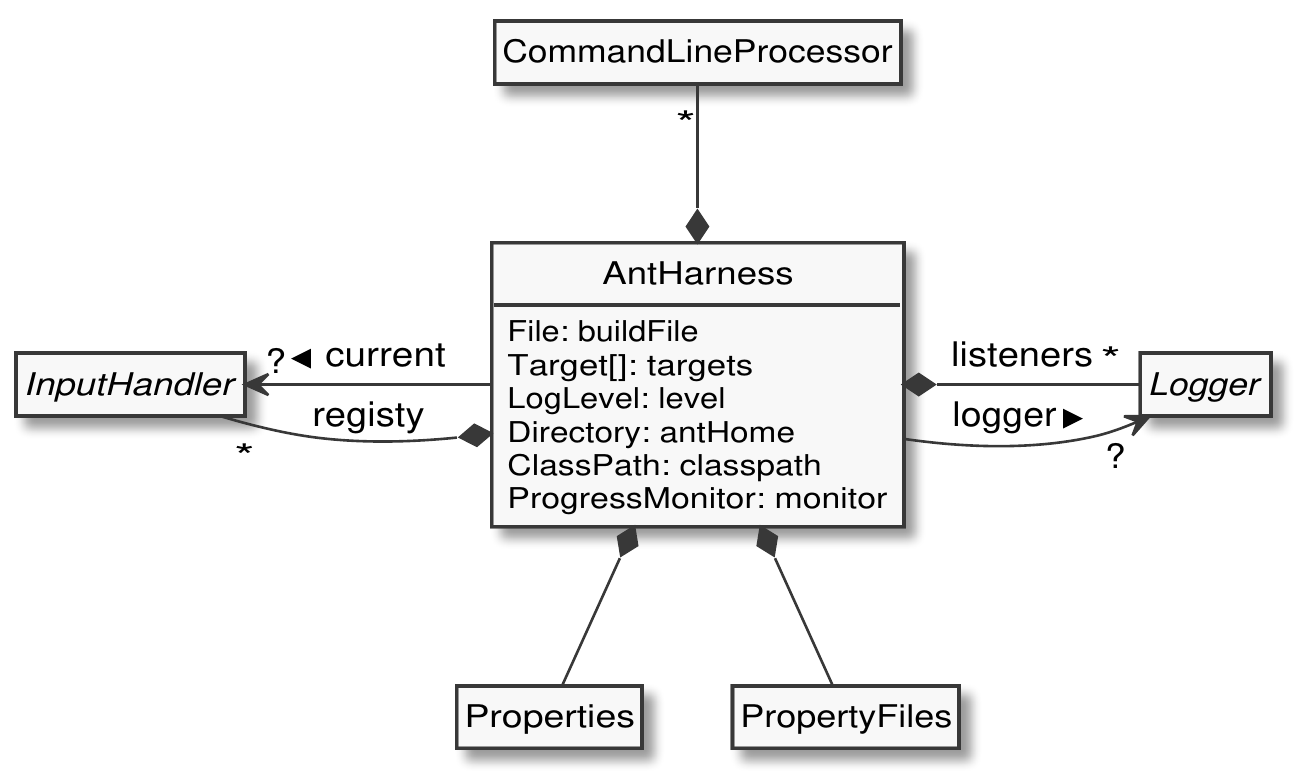}
\caption{A context diagram for the AntHarness.}
\label{ANT_CONTEXT}
\end{figure}

To ensure we can verify behaviour of components in the AntRunner, we have developed \href{https://JUnit.org/JUnit5/docs/current/user-guide/#extensions}{a JUnit5 Extension} called AntHarness that allows to create declarative tests that wrap around the AntRunner using configuration in Java annotations. Test classes with AntHarness annotations will run Ant build scripts inside AntRunner and make the configuration and runner available after execution has finished; An Eclipse developer can then use JUnit's assertion facilities to verify the outcomes. Since AntHarness is designed to run in the scope of \href{https://help.eclipse.org/latest/topic/org.eclipse.pde.doc.user/guide/tools/launchers/JUnit_launcher.htm}{Eclipse Plugin Tests}, all of Eclipse's debugging, logging and \href{https://wiki.eclipse.org/FAQ_How_do_I_use_the_platform_debug_tracing_facility}{tracing} support can be used with it, including attaching the debugger to the code of the Task under test or tracking an \href{https://www.eclipse.org/articles/Article-Progress-Monitors/article.html}{Eclipse Progress Monitor}. \autoref{ANT_CONTEXT} shows the AntRunner facilities the harness exposes in a context diagram.

% Add a code snippet of the Harness?

\subsection{User-level Testing using Ant Unit}
\label{ss:AntUnit}

Java developers will be familiar with JUnit 5. For them using the AntHarness is effective. But the real target audience for us is users of \ac{MDE} toolkits and workflows, not of the Java language. For them, testing must be easy and should only require a minimum of technological know-how. As we are designing this tooling for model-driven software developers, we think it is fair to assume that the users will have experience with some \href{https://en.wikipedia.org/wiki/XUnit}{xUnit}-style framework that uses naming conventions to find setups and tests and provides assertions to verify results.

Based on this, we integrated a solution to use Ant itself to perform integration tests. This allows to test build executions using Eclipse's built-in \href{https://help.eclipse.org/latest/topic/org.eclipse.platform.doc.user/reference/ref-anteditor.htm}{Ant editor and debugger facilities}, rather than the Java and Plugin Development environments. 

The facility we have chosen is provided by the \href{https://ant.apache.org/antlibs/antunit/}{AntUnit} framework. AntUnit transfers the ideas of JUnit to Ant: A regular Ant build script with conventionally named targets is executed dynamically in such a way that it prepares an environment, executes one or more test-case scripts, and validates the results. AntUnit can be combined with powerful model-driven testing facilities, such as \href{https://www.eclipse.org/epsilon/doc/eunit/}{Epsilon's EUnit}. This allows for combinatorial testing over models in the context of Ant-driven build flows.

\begin{lstlisting}[
language=Ant,
label=lst_antunit_call, 
caption=Invoking an AntUnit-based test from a build file.
]
    <au:antunit>
      <fileset dir="." includes="touch.xml"/>
      <au:plainlistener/>
    </au:antunit>
\end{lstlisting}

\begin{lstlisting}[
language=Ant,
label=lst_antunit_test, 
caption=An xUnit-like test expressed as an Ant Build file using AntUnit tasks.
]
<project xmlns:au="antlib:org.apache.ant.antunit">
  <!-- is called prior to the test -->
  <target name="setUp">
    <property name="foo" value="foo"/>
  </target>

  <!-- is called after the test, even if that caused an error -->
  <target name="tearDown">
    <delete file="${foo}" quiet="true"/>
  </target>

  <!-- the actual test case -->
  <target name="testTouchCreatesFile">
    <au:assertFileDoesntExist file="${foo}"/>
    <touch file="${foo}"/>
    <au:assertFileExists file="${foo}"/>
  </target>
</project>
\end{lstlisting}

\subsection{Installing Ant Workflow Features using P2}
\label{ss:P2}

The previous sections have covered how to execute workflows that use model-driven components of several stacks, but have not touched on how the various features required to produce a combined execution system need to be installed. In our approach, this task is performed by Eclipse's own installation system: The \href{https://help.eclipse.org/latest/topic/org.eclipse.platform.doc.isv/guide/p2_overview.htm}{\ac{P2}}. Inside the Eclipse workbench this tool is exposed as a set of user-interface widgets that \href{https://help.eclipse.org/latest/topic/org.eclipse.platform.doc.user/tasks/tasks-124.htm}{allow to install new software} from Eclipse installation websites. Eclipse features do not necessarily have to be related to the graphical \ac{UI}. In our case they are simply packaging for Ant \href{https://help.eclipse.org/latest/topic/org.eclipse.platform.doc.isv/reference/extension-points/org_eclipse_ant_core_antTasks.html}{Tasks}, \href{https://help.eclipse.org/latest/topic/org.eclipse.platform.doc.isv/reference/extension-points/org_eclipse_ant_core_antTypes.html}{Types} and other facilities.

The regular installation procedure for Ant tasks in features makes them immediately available in an Eclipse \ac{RCP} instance. In any Eclipse workbench a developer can install tasks and then start experimenting with model-driven-builds interactively.

There is a headless application whose functionality is equivalent to the graphical interactive installation in \ac{RCP} known as \href{https://help.eclipse.org/latest/topic/org.eclipse.platform.doc.isv/guide/p2_director.html?cp=2_0_20_2}{the director} which, provided with adequate parameters, \href{https://help.eclipse.org/latest/topic/org.eclipse.platform.doc.isv/guide/p2_director.html}{performs the same installation}. It behaves much like a package manager in the Unix operating system. It considers the installed base, the features the user intends to install, solves the dependency equation and provides an installation plan. It then creates a checkpoint and installs the new solution. 

\autoref{lst_eclipse_director} shows the command line invocation to install the set of plugins for C++ into an Eclipse instance in \lstinline{~/eclipse/}, downloading from the current release of the platform. 

\begin{lstlisting}[
language=Ant,
label=lst_eclipse_director, 
caption=Adding C++ Developer Tools to an Eclipse installation using the director.
]
eclipse \
  -application org.eclipse.equinox.p2.director \
  -repository http://download.eclipse.org/releases/latest/ \
  -installIU org.eclipse.cdt.feature.group \
  -destination ~/eclipse/ \
  -profile SDKProfile
\end{lstlisting}

\ac{P2} director also allows to update existing instances. So all that is required for a workflow with new requirements is the basic system and the locations of the update sites of the additional components.

The \ac{P2} director is not able to set any installation state, though. The \href{https://projects.eclipse.org/projects/tools.oomph}{Eclipse Oomph Installer} provides such functionality, but unfortunately, it is yet another tool integrated with the Eclipse \ac{UI} and \href{https://bugs.eclipse.org/bugs/show_bug.cgi?id=487626}{cannot be used in a headless environment} at this time.

\subsection{Creating portable Runtimes using Docker Containers}
\label{ss:Docker}

Installing Eclipse by hand on every system that needs to run workflows is a tedious process even with scripting provided by the director. It is also a source of inconsistencies and errors, as reproduction of the underlying environment is not easy and can have an effect due to native libraries and Java runtimes.

For this reason, we use Docker container images as lightweight encapsulations of existing installations. The build files that construct the container images - the \href{https://docs.docker.com/engine/reference/builder/}{Dockerfile}s - also allow us to control and change versions rather easily. Like with all other software artifacts, we use \ac{CI} to build these container images in a controlled fashion. 

The storage of these container images in repositories allows us to forward the installations of prepared workflow runtimes to other users almost instantly. Most importantly, keeping the workflows in container images allows us to use them effectively for \ac{CI}. 

Using pre-packaged container images does not limit the individualized use of the containers. Any user can \href{https://docs.docker.com/engine/reference/commandline/pull/}{obtain a pre-prepared container image}, \href{https://docs.docker.com/engine/reference/run/}{create a container and enter it}, call the director inside and change the setup in any way they care. They can also leave the container, \href{https://docs.docker.com/engine/reference/commandline/commit/}{commit the changes} and forward the new variant to someone else for use. Containers thus greatly increase the flexibility of the build process.

\subsection{Continuous Integration and \ac{DevOps} with Gitlab}
\label{ss:CI}

Model-driven workflows are often the first phase in a software construction process. They take in highly abstract data, and produce more language-specific and detailed software engineering artefacts. The toolchain that we have designed to drive our model-driven processes is specific to this purpose. It is \emph{not} made to compile C++, work with SAP/ABAP or MuleSoft RAML, or to render websites as HTML. While these things could be done using Ant as a general-purpose tool, this approach would constitute a \href{https://en.wikipedia.org/wiki/Law_of_the_instrument}{Golden Hammer} fallacy. Ready-made and highly adapted toolchains for all these technical environments are available in off-the-shelf container images which are maintained centrally and professionally.

Our obligation is that the toolchain \emph{activates} the corresponding build container images after the model-driven process has finished. We need a coordination layer to perform this and any following build steps. For this purpose we have chosen the Gitlab \ac{CI} system.

\subsubsection{Programming \ac{CI} using Multi-Technology Pipelines}
Gitlab expresses build processes in a declarative syntax in a file that uses shell commands that are valid in the containerized environment that the respective build phase runs in. At the start of each build pipeline, source code is retrieved based on a git commit hash and additional derived software artefacts are subsequently pumped through the various build phases of the pipeline using a transparent built-in file exchange mechanism alongside the source code.

\begin{figure}[ht]
\centering
\includegraphics[width=\textwidth]{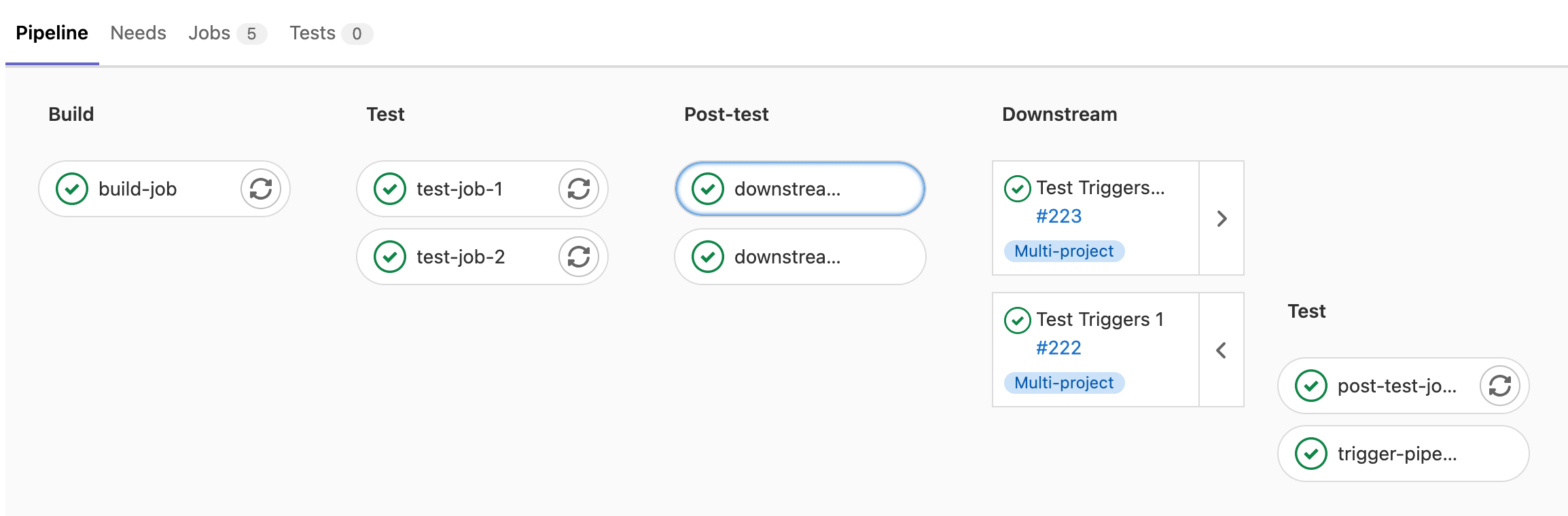}
\caption{A detail visualisation of a complex Gitlab pipeline showing phases and jobs.}
\label{fig:pipeline}
\end{figure}

Every time a commit is made, a pipeline is started, tests are executed and the developer receives feedback. \autoref{fig:pipeline} shows a complex multi-project pipeline visualisation. Execution time and computation cost are often significant, driving the design of build programs as first class artifacts and the segmentation of software into reusable artifacts. In this way, \ac{CI} and \ac{DevOps} drive software quality in a concrete, measurable fashion.

For our model-driven toolchains we invoke the AntRunner in a container installed with basic Eclipse and any model-driven tooling we require. The AntRunner executes the build file that is part of the source code checked out from the version control system. As a result, new source code files are created. As part of the Gitlab \ac{CI} build file we specify where these files are to be found. Gitlab will then take these files and move them into the next container.

Let us assume that we have generated some Java files and a Maven-specific Project Object Model (POM) build descriptor file. We now add a build phase in the Gitlab build descriptor that requires a container that is installed with the Maven build system. We start the Maven build system with a shell command and it reads the POM descriptor file. As a result, a Java application is now being built in a conventional way. Any Java developer - without any knowledge of model-driven processes - can understand what is happening here. As a result, we now have a ready-to-run Java application. We can now use the workflow to store the output in a binary repository. We can also add another phase with a container that holds tools used for deploying to a cloud provider. This can be proprietary, if desired, or based on a standard like Kubernetes or OpenShift.

This example showed how we performed \ac{CI} of a model-based artefact and followed it up with an immediate deployment.

\subsubsection{Reviews, A/B Tests, Beta Channels, Incidents}

The Gitlab \ac{CI} system also provides \href{https://docs.gitlab.com/ee/ci/variables/}{implicit parameters and variables} based on the build context that can be used to configure the \ac{CI} program in a declarative manner. These variables include the commit hash, commit ref, committer id and the locations of binary repositories associated with the build.

These implicit variables allow to integrate binary repositories into the process easily. Gitlab provides \href{https://docs.gitlab.com/ee/user/packages/package_registry/}{package repositories and registries for various technologies}. The build can upload binary artefacts to these package repositories.

Based on such parameters, it is now easy to deploy a version of the whole application for review: The pre-built binary that was produced and stored in the package registry in a previous phase can be retrieved from the registry and deployed. Developers can now review the application live and inspect the new feature.

With this setup many more \ac{DevOps} practices that seem advanced and theoretical become practically attainable: For example, variants of an application can be provided as part of \href{https://docs.gitlab.com/ee/development/experiment_guide/experimentation.html}{an A/B test}. \href{https://docs.gitlab.com/ee/operations/feature_flags.html}{Beta versions} can deployed for early adopters. An application's behaviour can be \href{https://docs.gitlab.com/ee/operations/}{be monitored} and feedback attached in the form of tickets.

The processes described in the previous paragraphs seems to imply that model-driven processes always have to be forward-engineering based. However, we could also consider the case where an existing application is read into model artefacts by using a reverse engineering system like Modisco to produce documentation of how the system can be understood. A project can serve as a log for discovery and experimentation. It is an effective vehicle for research, because it captures source state and all circumstances of a build, producing precise evidence. 

\subsubsection{Managing Technological Spaces with Sub-Projects}

\ac{MDE} processes excel at producing large, detailed code-bases. However, these days it is rare for projects of any significant size to use a single implementation language and technological platform. It is much more likely that specialised languages for the various aspects of a platform will be used in conjunction. The \ac{MDE} process will usually produce the software artefacts based on one set of consistent models in one directory, with derived artifacts for specific technologies residing in sub-directories.

However, in order to make use of the stereotypical builds for the various technology spaces it is useful to have projects that have a directory structure that conforms to the expectation of the language or technological space. Both developers and build processes in a technological space expect a git repository that contains the layout and commits for its technological concern, and nothing else. For example, the React front-end developer will not be interested that the model-driven system made a change to the underlying business model and persistence layers, if the server API stayed the same. From her point of view the git commit that announces that change is distracting noise in her git log.

For this reason, we will need to design the \ac{CI} system in such a way that it keeps the sub-projects for different technological spaces coordinated in a safe way, but at the same time keeps commits to them separated. In our environment we have achieved this by writing a set of specific Tasks that handle the creation of git sub-modules which map to individual Gitlab projects. The top-level project that references the other modules describes changes that happen at the top level. The individual projects are used by individual developers of the technologies. They behave like regular source-code-based projects.

Our company's existing solutions build full-stack applications based on several different technologies. Here, the sub-module management has direct application. In prior iterations, the build of these applications was complex and hence brittle. With the change to sub-modules this is much less complex.

As stated before, commits trigger builds in CI, so commits to the various sub-projects trigger corresponding builds in these sub-projects. This leads to the perception that sub-projects perform individual steps, when actually they are coordinated in a global way. Gitlab supports control of this by \href{https://docs.gitlab.com/ee/ci/pipelines/multi_project_pipelines.html#multi-project-pipeline-visualization}{tracking and visualizing the relationship between builds}.

This approach needs and leads to the stringent use of versioning. The various sub-projects generally do not use the latest version of another sub-project, but the latest verified version. This helps with stability and exposes API as a separate factor. It also helps to speed up selective builds of the sub-projects.

The management of versioning and tagging that is orchestrated by the sub-module tooling enables the use of binaries instead of souce code all the way down. With this approach results of a JavaScript build are stored in an NPM repository associated with the front-end project; Java project components will be deposited into a Maven repository associated with the Middleware project; and SQL packages will be stored within a general-purpose repository associated with the backend project. The Docker container images for these projects will be in one or more docker registries, and the Kubernetes deployment descriptor that references them will be in the helm project's registry.

With this setup, all an operations engineer needs from the development group to deploy the complete modelled server application is the address of the helm chart that describes the overall system. The knowledge required for deployment has been encoded by the developers as a helm chart program. In fact, the operations engineer in advanced process probably does not exist. If a new helm chart is committed, \ac{CD} occurs.

\subsubsection{Getting Interactive Input for the Workflow}

In the previous sections we have focused on the use of our architecture to perform builds based exclusively on git committed file input. However, in \ac{MDE} there are a lot of cases, where processes can benefit from input from a user. This is reflected in the fact that most Model-driven kits like \href{https://www.eclipse.org/epsilon/doc/eol/#context-independent-user-input}{Epsilon have synchronous input facilities}.

After all, domain expertise is one of the most valuable inputs into model-driven processes. Forcing a domain expert to learn the technicalities of a \ac{CI}/\ac{CD} system only to interact and provide information into a model-driven toolchain would imply very poor appreciation of the experts value. Engagement with the technology would likely be low.

Even current development approaches do not require software engineers to learn details of interaction with the \ac{CI}/\ac{CD} servers and services. Instead today we expect that build processes behave conversationally with those involved in what has become known as Chat-Ops. We expect that in our day-to-day chat interfaces the build process behaves like a participant. For example, we expect the build server to talk to us on a messaging channel (such as Mattermost or Slack) and inform us that a build has started, a new build phase has been entered, a bug has been found, or an artefact has been produced. The messages would depend on the needs of the communication participants. Likewise, we would expect the build system to also be able to receive feedback from the participants in the communication channel. For example, we would expect that the system would take note of a new development issue to be pursued by the team. We would expected it to respond to a team member typing something like this: \lstinline{"/issue Enlarge the font on the home screen."} asking questions pertinent to complete the capture of the issue.

In our company we have taken this idea of Chat-Ops one step further and created Model-Ops. In this case an \href{https://ant.apache.org/manual/inputhandler.html}{input facility provided by Ant} is connected to an arbitrary textual chat system. The system asks questions formatted in markdown syntax with references to domain items, and the build continues when it has captured a response from the team. The questions can be quite rich, containing full-format text, cross-references, tables, images and anything else you would expect in a post.

For example a reverse-engineering toolchain could ask something like: \lstinline{"Is HybridAdapterExtension.java a Java Bean? [y/n]"}, while hyperlinking the file to the git repository browser. The answer fills the knowledge gap and produces the correct data. The requesting process can now capture this information for later use.

Given that this interface can be attached to RCS/SMS text messages, Facebook messenger, Slack, Mattermost, IRC or Email it has a lot of reach for very little implementation effort. This approach can also be used in an asynchronous fashion where facilities ask a question in advance to return to a particular part of work once an answer becomes available.

\subsection{Using Kubernetes for Scaling and Online Service}
\label{ss:Kubernetes}

In our specific setup Gitlab and all associated container deployments are already hosted in Kubernetes on the cloud in Azure, and this covers the regular \ac{DevOps} requirements in terms of using \ac{MDE} workflows in a pipeline. However, Kubernetes also offers separate opportunities to expand and enhance the functionality of model-driven toolchains. As part of the following section we describe other prospects that Kubernetes can offer to \ac{MDE}.

\section{Outlook and Future Work}

While our current work is geared towards establishing a basis for \ac{MDE} workflows to harness model-driven processes as software engineering tools, the real interest is in exploring how we \emph{offer} model-driven facilities to others. There are three areas that we are interested in: How online-applications that are offering flexible model-driven services to users can be constructed (\autoref{ss:cloud_mde}), how we can support the construction of tools that use models to simplify software implementation and maintenance (\autoref{ss:botlearn}), and how modelling can be used to support \ac{DevOps} itself (\autoref{ss:modelops}).

\subsection{Applying Kubernetes for Cloud-based \ac{MDE}}
\label{ss:cloud_mde}

Instead of deploying a separate service execution for every run of a model-driven workflow, the workflow could be embedded into a persistently running server as part of an online service. Such services could then back interactive IDE's hosted on the cloud. We have noted previously that model-driven technologies are often employing reflective and interpreted languages and are consequently quite slow. For an interactive application this would be an impediment to adoption.

The usual response to slow interpretation is to devise a compiler to create program variations that perform the same task, but faster. Given that we are reusing existing \ac{MDE} components in our workflow that we are not maintaining ourselves, this is not an option. The only other avenue to increasing performance is parallelisation. Parallelisation in our approach can be achieved at different levels: At the level of the workflow by defining sections that can run in parallel, or at the level of whole build executions. The first option is again affected by the design of most Task implementations that we will be reusing: They are generally not thread-safe.

This leaves us with the option of performing parallelisation on the level of whole build process executions. For this purpose, we will aim to combine a variant of the \href{https://ant.apache.org/manual/Tasks/ant.html}{Ant} and \href{https://ant.apache.org/manual/Tasks/subant.html}{Subant} Tasks with the \href{https://github.com/kubernetes-client/java/}{Kubernetes Java Client}. This will allow a build to spawn other executions of the build as \href{https://kubernetes.io/docs/concepts/workloads/controllers/job/}{Kubernetes Jobs} to run in parallel and to balance the load where this desired.

As we distribute the workloads, access to resources becomes an issue. In this scenario, we do not have Gitlab's implicit file provision any more. The spawned jobs may require access to certain resources that live in the file system of the container image that spawned the new job. While there are good off-the-shelf solutions available for synchronising files and databases, sharing access to models in an organised manner is more challenging. We would assume that we would eventually implement tasks for our workflows that provide use of \href{https://help.eclipse.org/latest/topic/org.eclipse.emf.cdo.doc/html/Overview.html}{Connected Data Objects (CDO)} for versioned and branched target data, or \href{https://www.eclipse.org/hawk/}{Hawk} for file-based persistency, graph data or other clients.

Viewed from a greater distance, such application assemblies that operate on models effectively move modelling from our local environments into the cloud. They constitute real 'Cloud Modelling'. We intend to focus some of our research in this area to see what patterns are required to reuse existing components to assemble model-driven cloud-based applications.

\subsection{Teaching Bots to Code}
\label{ss:botlearn}

Evolved assistance in the creation of source code is getting traction in practice. The \href{https://copilot.github.com/}{Copilot} and \href{https://www.tabnine.com/}{Tabnine} services foreshadow that context-aware coding support will be a normality in the future. Given that much of the creation of source code generating \ac{MDE} templates is fairly repetitive, it should be feasible to create solutions that perform the bulk of abstraction work in providing coding support. Combined with the infrastructure for \ac{DevOps} and Model-Ops that BB8 offers and the prospect of parallelising work described in the previous section, an online service that supports a modelling engineer could be devised. We are currently focusing our research in this area.

\subsection{Using \ac{MDE} to Support \ac{DevOps}}
\label{ss:modelops}

\ac{DevOps} is the practice of encoding processes for software development and infrastructure. The flexibility that is required implies that the programming of these toolchains offers control of operational details. However, as these \ac{DevOps} programs evolve, they encounter the same issues as any other software. Gitlab for example began its \ac{CI} language as a simple YAML specification file. Today a dedicated parser manages \href{https://docs.gitlab.com/ee/ci/yaml/includes.html}{inclusions}, \href{https://docs.gitlab.com/ee/ci/yaml/#extends}{extensions}, \href{https://docs.gitlab.com/ee/ci/yaml/#stage-pre}{Around-advice}, \href{https://docs.gitlab.com/ee/ci/yaml/#needs}{dependency relationships}, \href{https://docs.gitlab.com/ee/ci/yaml/#parallelmatrix}{parallel matrix operations} and many other features. The resulting complexity is so substantial that an \href{https://docs.gitlab.com/ee/ci/interactive_web_terminal/}{interactive debugging shell} is now part of the system. With complexity and scale, models are becoming interesting to \ac{DevOps} engineers. Our company is looking at how the abstracting power of models can be used to support \ac{DevOps}.

\section{Conclusion}

Building an \ac{MDE} system that supports Agile practices and allows implementation of \ac{DevOps} is possible and provides a sound functional base for industrial application. Our implementation includes means for testing, installation, flexible deployment and workflow coordination, including limited interactivity, while considering the roles of stakeholders in the process. While our design is limited by the implications of \ac{EMF} as an Eclipse component, even the basic implementation provides a surprising amount of latitude.

We found that via aligning to the Ant build system we can combine many of the current MDE tools found in the Eclipse ecosystem. Even though at times we had to work around some tools because they were not designed with DevOps in mind, we found some positive results with this approach. We encourage the MDE community to consider DevOps toolchains as one of the Grand Challenges of MDE adoption in industry.

The experience we have gained also raises new questions with regard to increasing speed and performance of flexible synchronous \ac{MDE} services, automation of aspects of the construction of \ac{MDE} tooling and the application of \ac{MDE} approaches to \ac{DevOps} languages as a subject.

\begin{appendices}

% \section{Section title of first appendix}\label{secA1}

% An appendix contains supplementary information that is not an essential part of the text itself but which may be helpful in providing a more comprehensive understanding of the research problem or it is information that is too cumbersome to be included in the body of the paper.




\end{appendices}

\begin{acronym}
\acro{MDE}{Model-Driven Engineering}
\acro{DevOps}{Development and Operations}
\acro{CI}{Continuous Integration}
\acro{CD}{Continuous Deployment}
\acro{EMF}{Eclipse Modelling Framework}
\acro{MOF}{Meta-Object Facility}
\acro{CORBA}{Common Object Request Broker Architecture}
\acro{RCP}{Rich Content Platform}
\acro{UI}{User Interface}
\acro{PDE}{Plugin Development Environment}
\acro{P2}{Provisioning System 2}

\end{acronym}

\bibliography{bib/sosym}

\begin{thebibliography}{6}
\providecommand{\natexlab}[1]{#1}
\providecommand{\url}[1]{{#1}}
\providecommand{\urlprefix}{URL }
\providecommand{\doi}[1]{\url{https://doi.org/#1}}
\providecommand{\eprint}[2][]{\url{#2}}
 \bibcommenthead

\bibitem[{Bucchiarone et~al(2020)Bucchiarone, Cabot, Paige, and
  Pierantonio}]{DBLP:journals/sosym/BucchiaroneCPP20}
Bucchiarone A, Cabot J, Paige RF, et~al (2020) Grand challenges in model-driven
  engineering: an analysis of the state of the research. Softw Syst Model
  19(1):5--13. \doi{10.1007/s10270-019-00773-6},
  \urlprefix\url{https://doi.org/10.1007/s10270-019-00773-6}

\bibitem[{Debois(2011)}]{DBLP:journals/usenix-login/Debois11}
Debois P (2011) Devops from a sysadmin perspective. login Usenix Mag 36(6).
  \urlprefix\url{https://www.usenix.org/publications/login/december-2011-volume-36-number-6/devops-sysadmin-perspective}

\bibitem[{Fowler and Foemmel(2005)}]{fowler05integration}
Fowler M, Foemmel M (2005) Continuous integration,
  http://www.martinfowler.com/articles/continuousintegration.html

\bibitem[{S{\'{a}}nchez et~al(2020)S{\'{a}}nchez, Kolovos, and
  Paige}]{DBLP:conf/models/0001KP20}
S{\'{a}}nchez B, Kolovos DS, Paige RF (2020) To build, or not to build:
  Modelflow, a build solution for {MDE} projects. In: Syriani E, Sahraoui HA,
  de~Lara J, et~al (eds) MoDELS '20: {ACM/IEEE} 23rd International Conference
  on Model Driven Engineering Languages and Systems, Virtual Event, Canada,
  18-23 October, 2020. {ACM}, pp 1--11, \doi{10.1145/3365438.3410942},
  \urlprefix\url{https://doi.org/10.1145/3365438.3410942}

\bibitem[{Straeten et~al(2008)Straeten, Mens, and
  Baelen}]{DBLP:conf/models/StraetenMB08}
Straeten RVD, Mens T, Baelen SV (2008) Challenges in model-driven software
  engineering. In: Chaudron MRV (ed) Models in Software Engineering, Workshops
  and Symposia at {MODELS} 2008, Toulouse, France, September 28 - October 3,
  2008. Reports and Revised Selected Papers, Lecture Notes in Computer Science,
  vol 5421. Springer, pp 35--47, \doi{10.1007/978-3-642-01648-6\_4},
  \urlprefix\url{https://doi.org/10.1007/978-3-642-01648-6\_4}

\bibitem[{Whittle et~al(2013)Whittle, Hutchinson, Rouncefield, Burden, and
  Heldal}]{DBLP:conf/models/WhittleHRBH13}
Whittle J, Hutchinson JE, Rouncefield M, et~al (2013) Industrial adoption of
  model-driven engineering: Are the tools really the problem? In: Moreira A,
  Sch{\"{a}}tz B, Gray J, et~al (eds) Model-Driven Engineering Languages and
  Systems - 16th International Conference, {MODELS} 2013, Miami, FL, USA,
  September 29 - October 4, 2013. Proceedings, Lecture Notes in Computer
  Science, vol 8107. Springer, pp 1--17, \doi{10.1007/978-3-642-41533-3\_1},
  \urlprefix\url{https://doi.org/10.1007/978-3-642-41533-3\_1}

\end{thebibliography}


\end{document}